\documentstyle{aipproc}

\renewcommand{\mathbf}[1]{\mbox{\boldmath$#1$}}

\begin{document}


\title{Symmetrizing The Symmetrization Postulate}

\author{Michael York\footnote{975 S. Eliseo Dr. \#9, Greenbrae, CA 94904, USA}}        
\maketitle

\begin{abstract}
Reasonable requirements of (a) physical invariance under particle permutation and (b) physical completeness of state descriptions\cite{York}, enable us to deduce a Symmetric Permutation Rule(SPR): that by taking care with our state descriptions, it is always possible to construct state vectors (or wave functions) that are purely symmetric under pure permutation for all particles, regardless of type distinguishability or spin. The conventional exchange antisymmetry for two identical half-integer spin particles is shown to be due to a subtle interdependence in the individual state descriptions arising from an inherent geometrical asymmetry. For three or more such particles, however, antisymmetrization of the state vector for all pairs simultaneously is shown to be impossible and the SPR makes observably different predictions, although the usual pairwise exclusion rules are maintained. The usual caveat of fermion antisymmetrization -- that composite integer spin particles (with fermionic consitituents) behave only approximately like bosons -- is no longer necessary.
\end{abstract}

\section{Terminology}       

First let me express my deep gratitude to the organizers for the opportunity to present this paper at this conference.

The usual terminology of fermion/boson equates the statistics of a particle with the symmetry or antisymmetry of the wave function for pairs of such particles. In this talk, I will show that this equation is, at best, insufficiently specific and, in some circumstances, erroneous. To avoid confusion, I will first define the terms I use.

\begin{description}

\item[Halfon]: Particle with half-integer spin.

\item[Fullon]: Particle with integer spin. 

\item[Fermion]: Particle which obeys Fermi-Dirac statistics. (Is {\em not} defined by antisymmetry of the wave function.)

\item[Boson]: Particle which obeys Bose-Einstein statistics. (Is {\em not} defined by symmetry of the wave function.)

\item[Generalized Exclusion Rule]: Pairs of identical particles, for which all other quantum numbers are the same, must have even composite spin. (Note: this is the same rule for both halfons and fullons.)

\item[Spin-Statistics Theorem]: Identical halfons are fermions. Identical fullons are bosons. (Consequence of the Generalized Exclusion Rule.)

\end{description}

The chief reason for this terminology, breaking the equation between fermion statistics and antisymmetry, is that it is possible to obtain the same observable fermionic behavior for halfons with wave functions that are pure permutation symmetric. However, to see this requires a more thorough analysis of state descriptions and uniqueness of state vectors than is usual.

We usually label state vectors by a set of variables which we shall call a state description. Typically, state descriptions are lists of quantum numbers. Physical transformations that change these quantum numbers therefore change one state vector into another. Conventionally we assume we can choose a unique state vector for a given set of quantum numbers from an infinite set mutually related by arbitrary phase factors.

However, there are some physically significant transformations that leave quantum numbers unchanged but nevertheless change state vectors, even if only by a phase. To distinguish such state vectors, and still choose them uniquely, we need state descriptions that carry more information than just the quantum numbers.

\begin{description}

\item[Physical Completeness (Of A State Description)]: Requires the description of a state with sufficient precision to distinguish it from another state description related by any physically significant transformation -- even if quantum numbers are unchanged.

\item[Uniqueness Principle]: To choose a unique state vector for a given state description, the state description must be {\em physically complete}. (Follows from the definition above.)

\end{description}

The concept of {\em physical completeness} helps us to differentiate state vectors related by physically significant transformations that change only the phase. Since nature does not care about the order in which we describe individual particles in a multi-particle state, we can also write:

\begin{description}
\item[Permutation Invariance Principle]: Permutation of individual particles in a multi-particle state description is not a physically significant transformation.

\item[Symmetric Permutation Rule (SPR)]: For any multi-particle state for which each particle has a physically complete state description, each of which is independent of all the other individual state descriptions or their order, it is always possible to choose a state vector that is unchanged by (i.e. symmetric under) {\em pure permutation}, regardless of particle spin or even identity. (Follows from {\em physical completeness} and {\em permutation invariance}).

\item[Exchange Asymmetry]: When order dependent state descriptions are used, permutation implies a reversal of the order dependence. This may introduce a physically significant ``exchange'' transformation in addition to pure permutation and may result in a change of sign of the state vector.
\end{description}

\section{Physical Completeness And Spin}

We have argued that physical completeness is not a trivial matter of listing quantum numbers. We now need to explore what it means in the case of particles with spin. First, however, a few more terms need to be defined:
 
\begin{description}

\item[Spin Quantization Frame (SQF)]: The frame of reference in which we measure the spin component. 

\item[Canonical Frame]: The frame of reference in which we measure the position or momentum.

\end{description}

In general, the relative orientation of the SQF to the canonical frame need not always be a null rotation. Even when the axes of both frames coincide, for example, it could be given by a $2\pi$ rotation about some arbitrary axis.

We shall now consider those physically significant tranformations that can change the phase of the state vector for a particle with spin. 

\subsection{Single Particle SQFs}
In the usual methodology\cite{Wigner}\cite{AnyRQ}:
\begin{equation}\label{eqn:wigdef}
|Q,\mathbf{p},s,m(\hat{\mathbf{n}})> = U(B(\mathbf{p}))\ |Q,\mathbf{0},s,m(\hat{\mathbf{n}})>
\end{equation}
$|Q,\mathbf{0},s,m(\hat{\mathbf{n}})>$ is a rest frame eigenstate of spin $s$ and component $m$ in the direction $\hat{\mathbf{n}}$. $Q$ represents all other intrinsic quantum numbers. $U(B(\mathbf{p}))$ is boost operator for $\mathbf{0}\rightarrow \mathbf{p}$ (conventionally, but not necessarily, a {\em Lorentz} boost).
\begin{equation}\label{eqn:boostdef}
U(B(\mathbf{p}))\ = U(R(\hat{\mathbf{z}}\rightarrow\hat{\mathbf{p}}))U(B(p\hat{\mathbf{z}}))U(R^{-1}(\hat{\mathbf{z}}\rightarrow\hat{\mathbf{p}}))
\end{equation}
However, this is {\bfseries ambiguous} because $\hat{\mathbf{n}}$ does not uniquely specify the rotation $S\rightarrow C$ (S = SQF,C = canonical frame).

Instead, we prefer:
\begin{equation}\label{eqn:newwigdef}
|Q,\mathbf{p},s,m>^S = U(B(\mathbf{p}))\ |Q,\mathbf{0},s,m)>^S
\end{equation}
and, in terms of a standard base frame $B$,
\begin{eqnarray}\label{eqn:newwigdefBase}
|Q,\mathbf{p},s,m>^S  & = & |Q,\mathbf{p},s,m(R_{BS})>_B\nonumber\\
& = & \sum_{m'} D^s_{m' m}(R_{BS}) |Q,\mathbf{p},s,m'>^B
\end{eqnarray}
where $R_{BS}$ is the rotation $B\rightarrow S$. These latter two additions ($B,R_{BS}$) to the usual list of quantum numbers are clearly essential for a physically complete state description. 

Here are some commonly used examples:
\begin{eqnarray}\label{eqn:newwigdefexamples}
|Q,\mathbf{p},s,\lambda>^H & = & |Q,\mathbf{p},s,\lambda(N)>_H\\
|Q,\mathbf{p},s,m>^C & = & |Q,\mathbf{p},s,m(N)>_C\\
|Q,\mathbf{p},s,m>^C & = & |Q,\mathbf{p},s,m(R_{HC})>_H\nonumber\\
& = & \sum_{\lambda}\ D^s_{\lambda m}(R_{HC})\ |Q,\mathbf{p},s,\lambda>^H\\
|Q,\mathbf{p},s,\lambda>^H & = & |Q,\mathbf{p},s,\lambda(R_{CH})>_C\nonumber\\
& = & \sum_m\ D^s_{m \lambda}(R_{CH})\ |Q,\mathbf{p},s,m>^C
\end{eqnarray}
where $C$ is the canonical frame, $H$ is a helicity frame and $N$ is a null rotation. For massive particles, it is reasonable to choose $C$ for a base frame. To include massless particles, however, it is more reasonable to choose a frame $H$ for which the spin is quantized along the direction of motion. Of course, this still leaves the $x-$ and $y-$ axes undefined, but specifying the rotation $R_{CH}$ uniquely will resolve the matter.

\subsection{Two-Particle SQFs}
For two or more particles we must define $B$ for both particles. A common methodology is to choose $B = C$ for both particles and indeed it is quite possible to do so. However there is a very subtle complication that we must take care over: {\em an inherent geometrical asymmetry between any two vectors in a common frame of reference.}
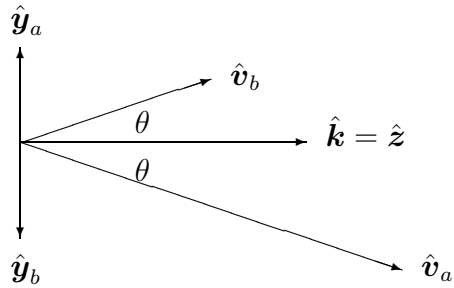
\begin{figure}[htbp]\begin{center}
\setlength{\unitlength}{1in}
\begin{picture}(4,1.8)
\put(1,1){\vector(0,1){0.5}}
\put(0.95,1.6){$\hat{\mathbf{y}}_a$}
\put(1,1){\vector(3,1){1.0}}
\put(2.1,1.3){$\hat{\mathbf{v}}_b$}
\put(1.6,1.05){$\theta$}
\put(1,1){\vector(3,0){1.5}}
\put(2.6,0.98){$\hat{\mathbf{k}}=\hat{\mathbf{z}}$}
\put(1.6,0.80){$\theta$}
\put(1,1){\vector(3,-1){2.0}}
\put(3.1,0.3){$\hat{\mathbf{v}}_a$}
\put(1,1){\vector(0,-1){0.5}}
\put(0.95,0.3){$\hat{\mathbf{y}}_b$}
\end{picture}
\end{center}
\end{figure}

As shown in the figure, we can select a common z-axis symmetrically by choosing $\mathbf{k}$ which bisects the two vectors $\hat{\mathbf{v}}_a$ and $\hat{\mathbf{v}}_b$. But as soon as we get to the other axes we find an asymmetry. For example, each of the two choices of y-axis shown, and their accompanying x-axis, is asymmetric with respect to $\hat{\mathbf{v}}_a$ and $\hat{\mathbf{v}}_b$. A classic manifestation of this asymmetry lies in the relationship between the polar angles of two momentum vectors in their CM frame. Although $\theta_a$ and $\theta_b$ are symmetrically related by $\theta_b = \pi - \theta_a$, the relationship between $\phi_a$ and $\phi_b$ is asymmetric: e.g. $\phi_b = \pi + \phi_a$.

To restore and ensure symmetry, when defining two-particle SQFs, it is better to use a common method to define $B$ {\em independently} for each particle. If we then rotate each separately to a common SQF when required, we can do so in a way that makes the asymmetry explicit. An helicity frame provides a good choice for this -- as long as we specify all axes using a symmetric method.

For ``current'' particle $c$, where $c = a$ or $b$, and ``other'' particle $o$, we define the independent helicity frames by
\begin{eqnarray}\label{eqn:helframe}
\hat{\mathbf{z}}_c & = & \hat{\mathbf{p}}_c\nonumber\\
\hat{\mathbf{y}}_c & = & \hat{\mathbf{p}}_c\times\hat{\mathbf{p}}_o / |\hat{\mathbf{p}}_c\times\hat{\mathbf{p}}_o|
\end{eqnarray}

To get to the canonical frame, we then have:
\begin{eqnarray}\label{eqn:heltocan}
\lefteqn{|Q_a,\mathbf{p}_a,s_a,m_a;Q_b,\mathbf{p}_b,s_b,m_b>^C}\nonumber\\
& = &  |(Q_a,\mathbf{p}_a,s_a,m_a(R_a))_{H_a};(Q_b,\mathbf{p}_b,s_b,m_b(R_b))_{H_b}>\nonumber\\
& = &  \sum_{\lambda_a \lambda_b}\ D^{s_a}_{\lambda_a m_a}(R_a)D^{s_b}_{\lambda_b m_b}(R_b)\ |Q_a,\mathbf{p}_a,s_a,\lambda_a;Q_b,\mathbf{p}_b,s_b,\lambda_b>^H
\end{eqnarray}

Since the helicity state vector on the right uses order independent and physically complete state descriptions, the SPR tells us that it is permutation symmetric. As long as $R_a,R_b$ are both uniquely specified and order independent then the canonical state vector on the left will also be permutation symmetric.

However, we can also write
\begin{eqnarray}\label{eqn:relrot}
R_b = R_a.R_{ba} = R_a.R_{\mathbf{k}}(\pm \pi)
\end{eqnarray}
where $R_{ba}$ takes the helicity frame of $b$ into that of $a$ and is a rotation by $\pm\pi$ about $\mathbf{k}$. The sign ambiguity encapsulates the problem with physical completeness for the two-particle state vector in the canonical frame, since, for a halfon, the difference between the two possible choices causes a sign difference between the two independent canonical state vectors that result from eqn. \ref{eqn:heltocan} for a given $R_a$.

If the state description does not specify unique choices of $R_a,R_b$ or, equivalently, $R_a,R_{ba}$, then it will not be complete. We would then not be able to determine how the state vector transforms under permutation. This ambiguity persists even in the limit that the momenta coincide and is inherent in the conventional way we define state vectors by simple lists of quantum numbers.

One alternative to choosing a fixed value of $R_{ba}$ is to fix the relative orientation between particle ``1'' and particle ``2'':
\begin{eqnarray}
R_2 = R_1.R_{21} = R_1.R_{\mathbf{k}}(\pm \pi)
\end{eqnarray}
This gives an order dependent method of specifying physical completeness:
\begin{eqnarray}
\lefteqn{|(Q_a,\mathbf{p}_a,s_a,m_a)^1;(Q_b,\mathbf{p}_b,s_b,m_b)^2>^C}\nonumber\\
& = &  |(Q_a,\mathbf{p}_a,s_a,m_a(R_a))_{H_a};(Q_b,\mathbf{p}_b,s_b,m_b(R_a.R_{21}))_{H_b}>\nonumber\\
\lefteqn{|(Q_b,\mathbf{p}_b,s_b,m_b)^1;(Q_a,\mathbf{p}_a,s_a,m_a)^2>^C}\nonumber\\
& = &  |(Q_b,\mathbf{p}_b,s_b,m_b(R_b))_{H_b};(Q_a,\mathbf{p}_a,s_a,m_a(R_b.R_{21}))_{H_a}>
\end{eqnarray}
and we see that there are two cases which tells us how to compute the exchange phase. The first case is:
\begin{eqnarray}
\lefteqn{R_b = R_a.R_{21}}\nonumber\\
\lefteqn{|(Q_b,\mathbf{p}_b,s_b,m_b)^1;(Q_a,\mathbf{p}_a,s_a,m_a)^2>^C}\nonumber\\
& = & (-)^{2s_a}|(Q_a,\mathbf{p}_a,s_a,m_a)^1;(Q_b,\mathbf{p}_b,s_b,m_b)^2>^C
\end{eqnarray}
and the second case is:
\begin{eqnarray}
\lefteqn{R_a = R_b.R_{21}}\nonumber\\
\lefteqn{|(Q_b,\mathbf{p}_b,s_b,m_b)^1;(Q_a,\mathbf{p}_a,s_a,m_a)^2>^C}\nonumber\\
& = & (-)^{2s_b}|(Q_a,\mathbf{p}_a,s_a,m_a)^1;(Q_b,\mathbf{p}_b,s_b,m_b)^2>^C
\end{eqnarray}

In each case re-ordering has forced a rotation by $2\pi$ on one particle's SQF (the two cases differing by which particle's has been rotated and therefore the two versions of the exchanged state vector differ by a $2\pi$ rotation to both particles' SQFs). This is a clear example of the relationship: {\em Exchange = Permutation + Physically Significant Transformation}; the physically significant transformation being a rotation.

The conventional description corresponds to the order dependent case for a very simple reason; because the conventional description is not physically complete, the exchange phase is indeterminate, unless we associate the missing information with the only free variable available -- the particle order.

\section{Composite Spin And Observable Effects}

As we previously noted, however, the observable effects of permutation invariance for identical particle pairs are uniquely described not by wave function symmetry or antisymmetry but by the generalized exclusion rules relating to composite spin.

To compute eigenstates of composite spin, we need a common SQF for both particles. Again, we must either specifically include the information concerning $R_{ba}$ or else assume an order dependent $R_{21}$. Clearly the conventional description corresponds to the latter case and we can prove the usual generalized exclusion rule for identical particles in the usual way. However, the ``antisymmetry'' usually associated with even composite spin for identical fermions is not real but is a {\em pseudo}-antisymmetrization due to a hidden order dependence in the {\em relative orientation} of the SQFs  --  even when their axes coincide. 

If we had used order independent SQFs, then the scalar coefficients would be symmetric for even composite spin, since the antisymmetric Clebsch-Gordon coefficients would be accompanied by phase factors differing by a sign, due to the difference between $R_{ba}$ and $R_{ab} ( = R_{ba}^{-1})$.

It should now be clear that the observed exclusion rules correspond to the effect of the asymmetric quality of eqn. \ref{eqn:relrot} on eigenstates of composite spin. In particular, any given pair in a multi-particle state will obey the usual exclusion rules. In the conventional approach, which equates fermion exclusion with antisymmetry, this is usually described as requiring ``complete'' antisymmetrization, by which is meant the simultaneous antisymmetrization with respect to all possible pairwise exchanges. There are, of course, $N(N-1)/2$ such independent pairs for $N$ identical halfons.

However, since it is composite spin that actually matters, rather than the antisymmetry, we must look, instead at the allowed spin combinations. The maximum number of independent simultaneous composite spin eigenstates of at least two particles is only $N-1$. Contrary to the usual description, we can therefore have at most $N-1$ simultaneous pseudo-antisymmetrizations of our $N$ halfon state. (State vectors for different combinations of these $N-1$ pairs will be related to each other by possible sign changes.) The allowed composite eigenstates will differ from those predicted by complete antisymmetrization\cite{York2}. Hence the difference between the SPR and the conventional complete antisymmetrization should be experimentally testable.

One very simple example of this lies with states of pairs of composite fullons. If these fullons are made up from constituent fermions then the conventional rule actually forbids these fullons from being true bosons. The usual explanation of this contradiction with experiment is that composite fullons are only {\em approximate} bosons and that when their wave functions overlap, their constituent fermions must be in excited states. The SPR however, predicts that fermionic or bosonic behavior is purely a quality of the overall spin of a system regardless of its constituents. Hence all fullons, elementary or composite, are exact bosons.

We now no longer need to think of fermions and bosons as different types of particle. The different statistics can be seen to originate purely in the difference in the way fullons and halfons combine to give eigenstates of composite spin.

\appendix
\section{Impossibility Of Complete Antisymmetrization}
I shall provide here a more rigorous proof that complete (simultaneous) antisymmetrization for all halfon pairs in states of three or more identical halfons is not possible.

First of all note that a general phase relation for state vectors that incorporate possible order dependence can be written in terms of a state vector that conforms to the SPR:
\begin{eqnarray}\label{eqn:arbitraryranking}
\lefteqn{|...Q_i,\mathbf{p}_i,s_i,m_i;Q_j,\mathbf{p}_j,s_j,m_j;Q_k,\mathbf{p}_k,s_k,m_k;...>^C}\\
& = & (-)^{...2n_is_i+2n_js_j+2n_ks_k...} |...(Q_i,\mathbf{p}_i,s_i,m_i)^0;
(Q_j,\mathbf{p}_j,s_j,m_j)^0;(Q_k,\mathbf{p}_k,s_k,m_k)^0;...>^C\nonumber
\end{eqnarray}
Superscript $0$ means that a unique order independent methodology has been used to define the rotation which takes the independent helicity frame of particle $i$ into the canonical frame, the phase factors for each particle $i$ arising from a possible $2\pi$ rotation of the order dependent state description relative to the order independent state description. Therefore, under any interchange, or combination of interchanges, the state vector will undergo a possible sign change 
\begin{eqnarray}\label{eqn:indivphasechange}
 (-)^{\sum_i 2N_i(c\rightarrow c') s_i} = (-)^{\sum_i 2(n_i' - n_i) s_i}
\end{eqnarray}
where $c$ represents the initial ordering, $c'$ represents the ordering brought about by the combination of interchanges and $n_i'$ are the new values of $n_i$ brought about by possible $2\pi$ rotations on the individual SQFs for each $i$ caused by this combination of interchanges.

Now, we note that the conventional symmetrization/antisymmetrization under pair exchange requires {\em non-interference} of any additional particles which may be present. That is, for any exchange $i\leftrightarrow j$, particle $k \neq i,j$ cannot be involved in determining the exchange sign. Hence $n_k' = n_k$ and, if $s_i = s_j = s$, then the exchange sign is $(-)^{2(n_i' - n_i + n_j' - n_j)s}$.

Now assume that all particles are identical halfons: $s_i = s$ for all $i$ and where $2s$ is odd. Then the phase relation must take the form:
\begin{eqnarray}\label{eqn:arbitraryfermionranking}
\lefteqn{|...(Q,\mathbf{p}_i,s,m_i)^1;(Q,\mathbf{p}_j,s,m_j)^2; (Q,\mathbf{p}_k,s,m_k)^3;...>^C}\nonumber\\
& = & (-)^{...n^1_i+n^2_j+n^3_k...} |...(Q,\mathbf{p}_i,s,m_i)^0;(Q,\mathbf{p}_j,s,m_j)^0;(Q,\mathbf{p}_k,s,m_k)^0;...>^C
\end{eqnarray}
and $1\leftrightarrow 2$ gives a sign change $(-)^{n^1_i - n^2_i + n^2_j - n^1_j}$. Similar sign changes hold for $2\leftrightarrow 3$ and $3\leftrightarrow 1$.

In particular, because the particle ordered $1$ can be any of $i,j,k$ and similarly for particles ordered $2,3$, then all three pairs can be exchanged by, for example, $1\leftrightarrow 2$ and the conventional halfon antisymmetrization rule then requires the simultaneous satisfaction of all three of the following conditions:
\begin{equation}
\begin{array}{rcccl}
\mbox{either} & n^1_i - n^2_i & \mbox{or} & n^2_j - n^1_j & \mbox{must be odd but not both,} \nonumber\\
\mbox{either} & n^1_j - n^2_j & \mbox{or} & n^2_k - n^1_k & \mbox{must be odd but not both and} \nonumber\\
\mbox{either} & n^1_k - n^2_k & \mbox{or} & n^2_i - n^1_i & \mbox{must be odd but not both.} 
\end{array}
\end{equation}
But this is {\em impossible}; if $n^1_i - n^2_i$ is odd(even), then to satisfy the first and third conditions, $n^2_j - n^1_j$ and $n^2_k - n^1_k$ must both be even(odd), thereby violating the second condition.

\end{document}